\documentclass[aps]{revtex4}
\usepackage{graphicx}
\usepackage{amsmath}
\usepackage{amssymb}

\begin{document}

\title{On the Hoop conjecture in Einstein gravity coupled to nonlinear electrodynamics}

\author{K.K. Nandi}
\email{kamalnandi1952@yahoo.co.in}
\affiliation{Zel'dovich International Center for Astrophysics, Bashkir State Pedagogical University, 3A, October Revolution Street, Ufa 450008, RB, Russia}
\affiliation{Department of Physics \& Astronomy, Bashkir State University, 47A, Lenin Street, Sterlitamak 453103, RB, Russia}
\affiliation{High Energy and Cosmic Ray Research Center, University of North Bengal, Siliguri 734 013, WB, India}
\author{R.N. Izmailov}
\email{izmailov.ramil@gmail.com}
\affiliation{Zel'dovich International Center for Astrophysics, Bashkir State Pedagogical University, 3A, October Revolution Street, Ufa 450008, RB, Russia}
\author{G.M. Garipova}
\email{goldberg144@gmail.com}
\affiliation{Zel'dovich International Center for Astrophysics, Bashkir State Pedagogical University, 3A, October Revolution Street, Ufa 450008, RB, Russia}
\author{R.R. Volotskova}
\email{fmfizika15@mail.ru}
\affiliation{Salavat Industrial College, 27 Matrosova Boulevard, Salavat 453259, RB, Russia}
\author{A.A. Potapov}
\email{a.a.potapov@strbsu.ru}
\affiliation{Department of Physics \& Astronomy, Bashkir State University, 47A, Lenin Street, Sterlitamak 453103, RB, Russia}


\begin{abstract}

The famous hoop conjecture by Thorne has been claimed to be\ violated in curved spacetimes coupled to linear electrodynamics. Hod \cite{Hod:2018} has recently refuted this claim by clarifying the status and validity of the conjecture appropriately interpreting the gravitational mass parameter $M$. However, it turns out that partial violations of the conjecture might seemingly occur also in the well known regular curved spacetimes of gravity coupled to \textit{nonlinear electrodynamic}s. Using the interpretation of $M$ in a generic form accommodating nonlinear electrodynamic coupling, we illustrate a novel extension that the hoop conjecture is \textit{not} violated even in such curved spacetimes. We introduce a Hod function summarizing the hoop conjecture and find that it surprisingly encapsulates the transition regimes between "horizon and no horizon" across the critical values determined essentially by the concerned curved geometries.
\end{abstract}

\maketitle


\section{Introduction}
\label{intro}
In 1972, Kip S Thorne \cite{Thorne:1972,Misner:1973} introduced a mathematically elegant and influential conjecture, called the hoop conjecture, that is widely believed to reflect a fundamental aspect of classical general relativty. The conjecture asserts that a self-gravitating matter configuration of mass $M$ will form an engulfing horizon if its circumferential radius $R=C/2\pi$ is equal to (or less than) the corresponding Schwarzschild radius $2M$ (in units $G=1,c=1$). That is, the hoop conjecture states that \cite{Thorne:1972} a black hole horizon exists if
\begin{equation}
C\leq 4\pi M(R).
\end{equation}

This relation has been supported by several studies (see, e.g., \cite{Redmount:1983,Abrahams:1992,Hod:2015,Peng:2018,Hod:2019}). Nevertheless, there has been some intriguing claims in the literature, based on a na\"{\i}ve application of the above relation, that the famous hoop
conjecture can be violated in charged curved spacetimes coupled to linear eletrodynamics \cite{deLeon:1987,Bonnor:1983}. Hod \cite{Hod:2018} has recently refuted this claim by clarifying the status and validity of the conjecture suggesting that the mass parameter on the r.h.s of (1) be appropriately interpreted as the gravitational mass $M(R)$ contained \textit{within} the engulfing hoop of radius $R$ and not as the total (asymptotically measured) mass $M_{\infty}$ of the entire spacetime.

In this paper, we shall be concerned with three well known curved spacetimes of Einstein gravity coupled to \textit{nonlinear electrodynamics} that are exact, everywhere regular including at the origin, asymptotically flat with ADM mass $M_{\infty}$ and charge $Q$. It turns out that, with $M=M_{\infty}$, the hoop conjecture can be partially violated in those spacetimes as well. To show that this need not be the case, we shall consider two classes of Ay\'{o}n-Beato and Garc\'{\i}a (AG) spacetimes \cite{AyonBeato:1998,AyonBeato:1999} and the Bardeen spacetime \cite{AyonBeato:2000} and examine the validity of the hoop conjecture in the light of Hod's interpretation \cite{Hod:2018} taken in its generic form accommodating nonlinear electrical energy.

\section{Hod's interpretation}
\label{sec:2}
The energy outside a charged ball of radius $R$ is
\begin{equation}
E_{\text{elec}}(r>R) = \int_{R}^{\infty}T_{0}^{0}4\pi r^{2}dr = \frac{Q^{2}}{2R},
\end{equation}%
where $T_{0}^{0}(r>R) = \frac{Q^{2}}{8\pi r^{4}}$ is the electric energy density in linear electrodynamics. Thus, for a charged ball of radius $R$, electric charge $Q$, the gravitational mass contained within ($r\leq R$) of the ball is given by%
\begin{equation}
M(r\leq R)=M_{\infty }-\frac{Q^{2}}{2R},
\end{equation}%
and so, according to the interpretation by Hod \cite{Hod:2018}, this mass should be used in the hoop conjecture instead of the asymptotic mass $M_{\infty}$, then:
\begin{equation}
\frac{C(R)}{4\pi M(r\leq R)}\leq 1 \Rightarrow \text{black hole horizon.}
\end{equation}%
This relation was used to show that the curved spacetime coupled to linear electrodynamic in \cite{deLeon:1987,Bonnor:1983} actually obeys the hoop conjecture.

Since we are concerned in this paper with spacetimes coupled to nonlinear electrodynamics, we shall use the generic formula for gravitational mass integrating the corresponding $T_{0}^{0}$:
\begin{equation}
M(r\leq R) = M_{\infty}-E_{\text{elec}} = M_{\infty}-\int_{R}^{\infty}T_{0}^{0}4\pi r^{2}dr.
\end{equation}%
A peculiarity common to the three curved spacetimes considered below is that the asymptotic ADM mass $M_{\infty}$ is \textit{independent} of the charge parameter $Q$, supporting the original idea of Born and Infeld \cite{Born:1934} to use nonlinear electrodynamics for proving the electromagnetic nature of mass. Therefore, Hod's interpretation embodied in (5) entails that the gravitational mass $M$ is plainly divided between two electromagnetic masses, one asymptotic $M_{\infty}$ and the other $E_{\text{elec}}$. Despite the curved spacetime coupled to nonlinear electrodynamics, such a straightforward division of masses surprisingly works well as far as the conjecture is concerned, as we will see shortly. We shall use (5) to study the validity of hoop conjecture (4).

\section{Curved spacetimes coupled to nonlinear electrodynamics}
\label{sec:3}
The three spacetimes under consideration follow from the gravitational action $S$ with source of nonlinear electrodynamics%
\begin{equation}
S = \int\sqrt{-g}d^{4}x\left[\frac{1}{16\pi}R-\frac{1}{4\pi}\mathcal{L}(F)\right],
\end{equation}%
where $R$ is the Ricci scalar and $\mathcal{L}$ is a function of $F = F_{\mu\nu}F^{\mu\nu}$. We omit further details and come directly to the relevant solutions.

\textit{(a) Ay\'{o}n-Beato and Garc\'{\i}a class 1 spacetime (AG1)}

The asymptotically flat metric is given by \cite{AyonBeato:1998}
\begin{eqnarray}
d\tau ^{2} &=&-A(r)dt^{2}+\frac{1}{A(r)}dr^{2}+r^{2}(d\theta ^{2}+\sin^{2}\theta d\psi ^{2}), \\
A(r) &=&1-\frac{2Mr^{2}}{(r^{2}+Q^{2})^{3/2}}+\frac{Q^{2}r^{2}}{(r^{2}+Q^{2})^{2}},
\end{eqnarray}%
with the associated asymptotically vanishing electric field $E$ given by
\begin{equation}
E=Qr^{4}\left[ \frac{r^{2}-5Q^{2}}{(r^{2}+Q^{2})^{4}}+\frac{15}{2}\frac{M}{(r^{2}+Q^{2})^{7/2}}\right],
\end{equation}%
where the $M$ is the asymptotic ADM mass, hereinafter to be understood as $M_{\infty}$, and $Q$ is related to the electric charge. AG define two dimensionless parameters $s,x$ as%
\begin{equation}
s = \frac{Q}{2M_{\infty}},\quad x=\frac{r}{Q},
\end{equation}%
and by numerically solving two simultaneous equations%
\begin{equation}
A(x_{c},s_{c})=0\text{ and }\partial _{x}A(x_{c},s_{c})=0,
\end{equation}%
they find two critical values, $s_{c}$ and $x_{c}$, given by%
\begin{equation}
s_{c}=0.317, x_{c}=1.58,
\end{equation}%
Keeping $x_{c}=1.58$ fixed, the transition between "no horizon to black hole horizon" regime is marked by the critical parameter $s_{c}$ as follows \cite{AyonBeato:1998}:
\begin{eqnarray}
s &>& s_{c} \quad \Rightarrow \quad \text{no horizon} \\
s &<& s_{c} \quad \Rightarrow \quad \text{black hole horizon} \\
s &=& s_{c} \quad \Rightarrow \quad \text{two coincident horizons.}
\end{eqnarray}%
Let us look at Eq.(5). The electric energy density $T_{0}^{0}$ can be obtained from the Einstein equations $G_{\alpha\beta} = 8\pi T_{\alpha\beta}$, which yield \cite{Manko:2016}
\begin{eqnarray}
-T_{0}^{0}(r>R) &=& -\frac{1}{8\pi}G_{00}g^{00} \nonumber\\
&=& \frac{Q^{2}\left(r^{2}-3Q^{2}+6M_{\infty}\sqrt{r^{2}+Q^{2}}\right)}{8\pi \left(r^{2}+Q^{2}\right)^{3}}.
\end{eqnarray}%
Integrating, we find the energy outside a ball of radius $R$ to be
\begin{eqnarray}
&&E_{\text{elec}}(r >R)=\int_{R}^{\infty }T_{0}^{0}4\pi r^{2}dr = \frac{Q^{2}R^{3}}{2\left(R^{2}+Q^{2}\right)^{2}}  \nonumber \\
&& + \frac{2M_{\infty}\left[Q^{2}\sqrt{R^{2}+Q^{2}}+R^{2}\left(\sqrt{R^{2}+Q^{2}}-R\right)\right]} {2\left(R^{2}+Q^{2}\right)^{3/2}}.
\end{eqnarray}%
Therefore, Eq.(4) becomes%
\begin{eqnarray}
\frac{M(r\leq R)}{R} &=&\frac{M_{\infty }-E_{\text{elec}}}{R}  \notag \\
&=&\frac{2b-a^{2}\left( \sqrt{1+a^{2}}-2b\right) }{2\left( 1+a^{2}\right)^{5/2}},
\end{eqnarray}%
where we have used the dimensionless parameters $a,b$ defined by%
\begin{equation}
a=\frac{Q}{R}\text{, }b=\frac{M_{\infty}}{R}\Rightarrow s=\frac{a}{2b}.
\end{equation}

Taking into account Eq.(5), the hoop conjecture (4) yields what one may call the Hod function $H(b,s)$ for brevity:
\begin{eqnarray}
\frac{C(R)}{4\pi M(r\leq R)}&=&\frac{(1+4b^{2}s^{2})^{5/2}}{%
2b-4b^{2}s^{2}\left(\sqrt{1+4b^{2}s^{2}}-2b\right)} \nonumber \\
&\equiv& H_{1}(b,s),
\end{eqnarray}%
the subscript $1$ refers to solution AG1. Eq.(19) says that there are three variables connected by one equation $s=\frac{a}{2b}$, so we can choose two independent variations in $b$ and $s$. Since the transitions in (13-15) are described only in terms of $s_{c}$, so we need to vary $s$ through $s_{c}$, and $b$ through $b_{c}$ given by
\begin{eqnarray}
2b_{c} &=&\left. \frac{2M_{\infty }}{R}\right\vert _{c}=\left. \frac{Q}{R}%
\right\vert _{c}\times \left. \frac{2M_{\infty }}{Q}\right\vert _{c} \\
&=&\frac{1}{x_{c}}\times \frac{1}{s_{c}}=\frac{1}{1.58}\times \frac{1}{0.317}%
=1.99657.
\end{eqnarray}

Note that horizon properties of AG1 are described with fixed $x_{c}=1.58$, hence $b$ ($=\frac{1}{2sx_{c}}$) depends only on $s$. Now, take a value in the "no-horizon" range, say, $s=0.4$. To protect the hoop conjecture in this case with\ $M=M_{\infty }$, one would need to show that $\frac{C}{4\pi M_{\infty }}>1$. On the other hand, using (10) and (19), we find
\begin{equation}
\frac{C}{4\pi M_{\infty }}=\frac{1}{2b}=sx_{c}=0.632<1,
\end{equation}%
which\textit{\ }violates the conjecture. However, the other half of the story, viz., the horizon range $s\leq 0.317$ is consistent with the hoop conjecture $\frac{C}{4\pi M_{\infty }}=sx_{c}\leq 1$. So, overall, the use of $M=M_{\infty}$ \textit{partially violates} the hoop conjecture but this violation is only apparent. Using (4,5) we can restore the validity of the conjecture in the entire range of $s$. This argument can be applied to the remaining two solutions as well, hence will not be reproduced further.

\textit{(b) Ay\'{o}n-Beato and Garc\'{\i}a class 2 spacetime (AG2)}

The asymptotically flat metric is given by \cite{AyonBeato:1999}

\begin{eqnarray}
d\tau ^{2} &=&-A(r)dt^{2}+\frac{1}{A(r)}dr^{2}+r^{2}(d\theta ^{2}+\sin^{2}\theta d\psi ^{2}), \\
A(r) &=&1-\frac{2M}{r}\left[ 1-\tanh \left( \frac{Q^{2}}{2Mr}\right) \right],
\end{eqnarray}%
with the associated asymptotically vanishing electric field $E$ given by
\begin{eqnarray}
E&=&\frac{Q}{4Mr^{3}}\left[ 1-\tanh ^{2}\left( \frac{Q^{2}}{2Mr}\right)\right] \times \nonumber \\
&&\left[ 4Mr-Q^{2}\tanh\left( \frac{Q^{2}}{2Mr}\right) \right].
\end{eqnarray}%
Like before, the electric energy density is \cite{Manko:2016}
\begin{equation}
-T_{0}^{0}(r>R)=\frac{Q^{2}}{8\pi r^{4}}\sec h^{2}\left( \frac{Q^{2}}{%
2M_{\infty }r}\right) .
\end{equation}%
Integrating, we find the energy outside a ball of radius $R$ to be
\begin{eqnarray}
E_{\text{elec}}(r &>&R)=\int_{R}^{\infty }T_{0}^{0}4\pi r^{2}dr  \nonumber \\
&=& M_{\infty }\tanh \left( \frac{Q^{2}}{2M_{\infty }r}\right) .
\end{eqnarray}%
For AG2, the two dimensionless parameters are%
\begin{equation}
s = \frac{Q}{2M_{\infty }},x=\frac{2M_{\infty }r}{Q^{2}}
\end{equation}

Defining as before
\begin{equation*}
a=\frac{Q}{R},b=\frac{M_{\infty }}{R}\Rightarrow a=2bs
\end{equation*}%
The Hod function follows as%
\begin{equation}
H_{2}(b,s)=\frac{1}{2b\left[ 1-\tanh \left( 2bs^{2}\right) \right] },
\end{equation}%
the subscript $2$ refers to solution AG2. The critical values are
\begin{eqnarray}
s_{c} &=&0.53,x_{c}=1.56 \\
&\Rightarrow &2b_{c}=\left. \frac{2M_{\infty }}{R}\right\vert _{c}=\frac{1}{%
1.56}\times \frac{1}{(0.53)^{2}}=2.28204
\end{eqnarray}

\textit{(c) Bardeen spacetime}

he asymptotically flat metric is given by \cite{AyonBeato:2000}

\begin{eqnarray}
d\tau ^{2} &=&-A(r)dt^{2}+\frac{1}{A(r)}dr^{2}+r^{2}(d\theta ^{2}+\sin
^{2}\theta d\psi ^{2}), \\
A(r) &=&1-\frac{2Mr^{2}}{(r^{2}+Q^{2})^{3/2}},
\end{eqnarray}%
with the associated electromagnetic field tensor $F_{\mu \nu}$ given by
\begin{equation}
F_{\mu \nu }=2\delta _{\lbrack \mu }^{\theta }\delta _{\nu ]}^{\psi }Q\sin
\theta .
\end{equation}

The charge parameter $Q$ was originally interpreted as describing the electric charge, but later identified by Ay\'{o}n-Beato and Garc\'{\i}a \cite{Hod:2018} as representing a magnetic monopole coupled to nonlinear electrodynamics. Like before, the electric energy density is \cite{Manko:2016}
\begin{equation}
-T_{0}^{0}(r>R)=\frac{3M_{\infty }Q^{2}}{4\pi (r^{2}+Q^{2})^{5/2}}.
\end{equation}%
Integrating, we find the energy outside a ball of radius $R$ to be
\begin{eqnarray}
E_{\text{elec}}(r >R)&=&\int_{R}^{\infty }T_{0}^{0}4\pi r^{2}dr  \notag \\
&=&\frac{M_{\infty }\left( Q^{2}\sqrt{Q^{2}+R^{2}}+R^{2}\left( \sqrt{%
Q^{2}+R^{2}}-R\right) \right) }{\left( Q^{2}+R^{2}\right) ^{3/2}}.
\end{eqnarray}%
As in AG1, the two dimensionless parameters are%
\begin{equation}
s=\frac{Q}{2M_{\infty }},\text{ }x=\frac{r}{Q}
\end{equation}%
with their critical values%
\begin{equation}
s_{c}=\frac{2}{\sqrt{27}},x_{c}=\sqrt{2},b_{c}=\frac{\sqrt{27}}{4\sqrt{2}}%
=0.918559.
\end{equation}

Proceeding exactly as in (a), the Hod function follows as%
\begin{equation}
H_{3}(b,s)=\frac{\left( 1+4b^{2}s^{2}\right) ^{3/2}}{2b},
\end{equation}%
the subscript $3$ referring to the Bardeen solution \cite{AyonBeato:2000}.

The 3D plots of $H_{1},H_{2}$ and $H_{3}$, where $s$ and $b$ are varied through their critical values, are combined and shown in Fig.1. We have checked that the plots are somewhat insensitive to the variation of $b$. However, the combined 3D plot is not very transparent for reading out the transition points stated in (13-15). For clarity, we show in Fig.2 the plots of $H_{1},H_{2}$ and $H_{3}$, which are just a section of Fig.1 at some average value of $b$ around $b_{c}$, say $b=1$. Fig.2 excellently shows the transitions points between "no horizon and horizon" regimes corresponding to each solution in \textit{(a)-(c)}.

\begin{figure}[!ht]
  \centerline{\includegraphics[scale=1.25]{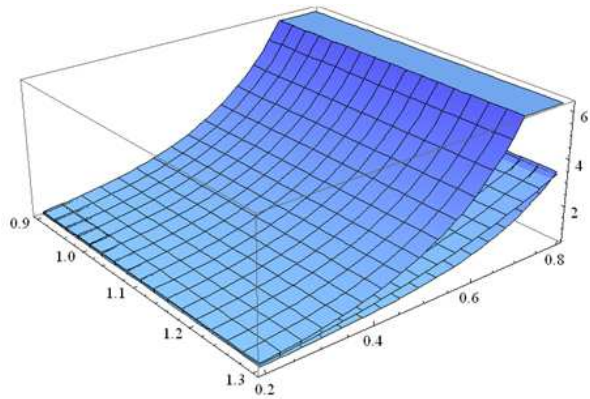}}
  \caption{The functions $H_{i}(b,s) $, $i=1,2,3$ are plotted within the interval $b\in \lbrack 0.9,1.3]$ and $s\in \lbrack 0.2,0.8]$. The intervals are so chosen as to ensure that the variables pass through the respective critical values.}
  \label{Veff}
\end{figure}

\begin{figure}[!ht]
  \centerline{\includegraphics[scale=1.2]{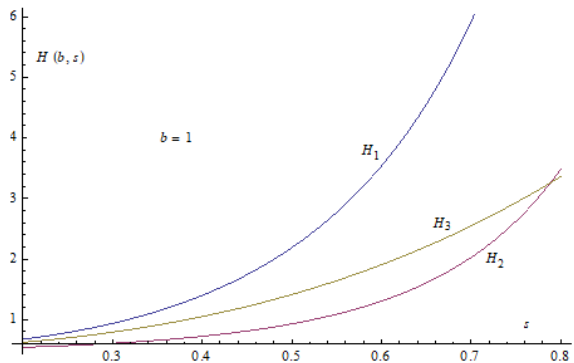}}
  \caption{For clarity, the functions $H_{i}(b,s)$, $i=1,2,3$ are plotted against $s$ at some fixed average value of $b$ around $b_{c}$, say $b=1$. For $H_{1}$, transition point in the metric AG1 is $s=s_{c}=0.317$. Similarly, for $H_{2}$, transition point in the metric AG2 is $s=s_{c}=0.53$ and for $H_{3}$, the transition point in the Bardeen metric is $s=s_{c}=\frac{2}{\protect\sqrt{27}}=0.3849$. The behavior of the curves clearly demonstrates that for $s\leq s_{c}$, corresponding to existence of "horizon (14,15)", we find $H_{i}\leq 1 $ and for\ $s>s_{c}$, corresponding to existence of "no horizon (13)", we find $H_{i}>1$, all nicely showing that the hoop conjecture is not violated.}
  \label{Veff}
\end{figure}

\section{Conclusions}
\label{sec:concl}
Hod \cite{Hod:2019} has recently shown that Thorne's hoop conjecture is not violated, despite claims to the contrary, in the charged curved spacetimes if the concerned gravitational mass $M$ \ is appropriately interpreted (we called it Hod's interpretation). Present paper is an extension of this result to curved spacetimes coupled to \textit{nonlinear electrodynamics}, where a\ na\"{\i}ve application of the conjecture\ using $M_{\infty }$ would also appear to lead to its \textit{partial} violation, as argued around (23). This need not be the case.

The novel result we obtained is that, despite electrodynamic nonlinearity, the hoop conjecture is \textit{not} violated in the spacetimes as exemplified in the text\textit{, }provided Hod's interpretation, embodied in Eq.(5),\ is taken with generic form of $E_{\text{elec}}$ accommodating the nonlinearity. The Hod function $H(b,s)$ we introduced in (20), (30) and (40) summarizes the conjecture. Fig.1 plots $H_{i}(b,s)$ in 3D for intervals of $\left( b,s\right) $ that contain their respective critical values (it may be verified that the 3D plots are actually almost insensitive to $b$). For clarity, we fix an average value around the critical values of $b$, say $b=1$, without any loss of rigor. It can be immediately seen that the curves in Fig.2 surprisingly encapsulate the transition between the "no horizon and horizon" regimes across the critical values $s_{c}$ determined \textit{purely} by the geometry.

\end{document}